\documentstyle[11pt,aaspp4]{article}

\begin{document}

\title{A Comparison of Boltzmann and Multigroup Flux-Limited Diffusion Neutrino Transport
During the Postbounce Shock Reheating Phase\\ in Core Collapse Supernovae}

\author{O. E. B. Messer\altaffilmark{1,2},
A. Mezzacappa\altaffilmark{1,2}, 
S. W. Bruenn\altaffilmark{3}, and 
M. W. Guidry\altaffilmark{1,2}
}
\vspace{5.0in}
\begin{center}
Submitted for publication in {\it The Astrophysical Journal}
\end{center}

\altaffiltext{1}{Physics Division, Oak Ridge National
Laboratory, Oak Ridge, TN 37831--6354}
\altaffiltext{2}{Department of Physics and Astronomy, University of Tennessee,
Knoxville, TN 37996--1200}
\altaffiltext{3}{Department of Physics, Florida Atlantic University, Boca Raton, FL 33431--0991}

\newpage

\begin{center}
{\large {\bf Abstract}}
\end{center}

We compare Newtonian three-flavor multigroup Boltzmann ({\small MGBT})
and (Bruenn's) multigroup 
flux-limited diffusion ({\small MGFLD}) neutrino transport in postbounce core collapse 
supernova environments. 
We focus our study on quantities 
central to the postbounce neutrino heating mechanism for reviving the 
stalled shock. Stationary--state three--flavor neutrino distributions 
are developed in thermally and hydrodynamically frozen time slices 
obtained from core collapse and bounce simulations that implement 
Lagrangian hydrodynamics and {\small MGFLD} neutrino transport. 
We obtain distributions for time slices at 106 ms and 233 ms after core bounce 
for the core of a 15 ${\rm M}_{\odot}$ progenitor, and at 156 ms 
after core bounce for a 25 ${\rm M}_{\odot}$ progenitor.   
For both transport methods, the electron neutrino and antineutrino 
luminosities, {\small RMS} energies, and mean inverse flux factors, all 
of which enter the neutrino heating rates, are computed as functions
of radius and compared. The net neutrino heating rates are also 
computed as functions of radius and compared. 

Notably, we find significant differences in neutrino luminosities and mean inverse 
flux factors between the two transport methods for both precollapse models and for 
all three time slices.  In each case, 
the luminosities for each transport method begin to diverge above the neutrinospheres, 
where the {\small MGBT} luminosities become larger than their {\small MGFLD} counterparts, finally
settling to a constant difference maintained to the edge of the core.    
We find that the mean inverse flux factors, which describe the degree of forward 
peaking in the neutrino radiation field,
also differ significantly between the two transport methods, with  
{\small MGBT} providing more isotropic radiation fields in the gain region.  

Most important, we find, for a region above the gain radius, net heating 
rates for {\small MGBT} that are as much as $\sim2$ times   
the corresponding {\small MGFLD} rates, 
 and net cooling rates 
below the gain radius that are typically $\sim$0.8 times the {\small MGFLD} rates. 
These differences stem from differences in the neutrino luminosities and mean inverse flux factors, 
which can be as much as 11\% and 24\%, respectively. 
They are greatest at earlier postbounce times for a given progenitor mass 
and, for a given postbounce time, 
 greater for greater progenitor mass.
We discuss the ramifications these new results 
have for the supernova mechanism.

\keywords{(stars:) supernovae: general}
\newpage

\section{\bf Introduction}

Ascertaining the core collapse supernova mechanism is a long-standing
problem in astrophysics. The current paradigm begins with the collapse of a 
massive star's iron core and the generation of an outwardly propagating shock 
wave that results from core rebound. Because of nuclear dissociation 
and neutrino losses, 
the shock stagnates. This sets the stage for a shock reheating mechanism 
whereby neutrino energy deposition via electron neutrino 
and antineutrino  
absorption on nucleons behind the shock reenergizes it (Bethe \& Wilson 1985; 
Wilson 1985). 

The shock reheating phase is essential to the supernova's success, but it 
is precisely this phase that is difficult to simulate realistically. During
shock reheating, core electron neutrinos and 
antineutrinos are radiated from their respective neutrinospheres, 
and a small fraction of this radiated energy is absorbed in the exterior 
shocked mantle. The shock reheating depends sensitively on the electron neutrino 
and antineutrino luminosities,  
spectra (best characterized by the {\small RMS} energies),
 and angular distributions in the region 
behind the shock (e.g., see Burrows \& Goshy 1993, Janka \& M\"{u}ller 1996, 
Mezzacappa et al. 1998b).  
These, in turn, depend on the neutrino transport in the 
semitransparent region encompassing the neutrinospheres, necessitating a 
neutrino transport treatment that is able to transit accurately and 
seamlessly between neutrino-thick and neutrino-thin regions. 

Various neutrino transport approximations have been implemented in simulating
core collapse supernovae.
The most sophisticated approximation, which
naturally has been used in detailed one-dimensional simulations, is multigroup 
flux-limited diffusion ({\small MGFLD}; e.g., Bowers \& Wilson 1982, Bruenn 1985, 
Myra et al. 1987). 
{\small MGFLD} closes the neutrino radiation hydrodynamics hierarchy of equations 
at the level of the first moment (the neutrino flux) by imposing a relationship
between the flux and the gradient of the neutrino energy density (the zeroth
moment).  For example,  

\begin{equation}
F_{\nu}=-\frac{c\Lambda}{3}\frac{\partial U_{\nu}}{\partial r}+...,
\label{eq:mgfld}
\end{equation}

\begin{equation}
\Lambda = \frac{1}{1/\lambda + |\partial U_{\nu}/\partial r|/3U_{\nu}},
\label{eq:lambda}
\end{equation}

\noindent where $\lambda$ is the neutrino mean free path, and $U_{\nu}$ and $F_{\nu}$ 
are the neutrino energy density and flux (Bruenn 1985).
 [Other forms for the flux-limiter $\Lambda$ can be found in Bowers \& Wilson (1982), 
Levermore \& Pomraning (1981), 
and Myra et al (1987).]
 Whereas the limits $\lambda \rightarrow 0$
  and $\lambda \rightarrow \infty$  produce
the correct diffusion and free streaming fluxes, it is in the critical intermediate 
regime where the {\small MGFLD} approximation is of unknown accuracy.  
Unfortunately, the quantities central 
to the postshock neutrino heating mentioned earlier are determined in this 
regime, and given the sensitivity of the neutrino heating to these quantities it becomes 
necessary to consider more accurate transport schemes.  Moreover, in detailed one-dimensional 
simulations that have implemented elaborate {\small MGFLD} neutrino transport (e.g., see
 Bruenn 1993, Wilson \& Mayle 1993, and Swesty \& Lattimer 1994), explosions were not 
obtained unless the neutrino heating was boosted by additional phenomena, such as convection. 
This leaves us with at least two possibilities to consider:  (1) Failures to produce 
explosions in the absence of additional phenomena, such as convection, have resulted 
from inexact neutrino transport.  (2) Additional phenomena may be essential 
in obtaining explosions.  

Option (1) requires further comment.  All investigators  agree convection will occur during the 
shock reheating, explosion initiation phase in core collapse supernovae (Herant, Benz, \&
 Colgate 1992; 
Miller et al. 1993; Herant et al. 1994; Burrows et al. 1995, Janka \& M\"{u}ller 1996, 
Mezzacappa et al. 1998b).  Therefore, 
strictly speaking, all investigators agree the flow will not be spherically symmetric. 
However, although convection will certainly occur, it may play no significant role in 
{\em initiating} the explosion.  It is with this 
distinction in mind that option (1) need be considered.  For example, significant 
neutrino-driven convection was seen in recent multidimensional simulations 
employing one-dimensional {\small MGFLD} neutrino transport (Mezzacappa et al. 1998b); 
however, the angle-averaged shock radius, among other quantities, did not differ significantly 
from its one-dimensional counterpart, and no explosion was obtained.  

Ultimately, any successful model of core collapse supernovae will have to 
reproduce observables that clearly do not originate from spherically symmetric explosions,
 the most obvious of which is 
neutron star kicks.  At this point, whether or 
not these kicks are generated during the initiation of the explosion or  
shortly thereafter is an open question.  Note in this regard 
that simulations that have invoked 
multidimensional effects such as convection to explain such kicks have had difficulty 
generating, for example,  
adequate kick velocities. Mechanisms invoking  
convection, or aspherical neutrino emission resulting from
convection, have not been able to produce kicks in excess of about 300
km/s, which therefore cannot account for the fastest pulsars --- for
example, PSR 2224+65, which has a velocity around 800 km/s --- 
(Janka \& M\"{u}ller 1994; Burrows \& Hayes 1995, 1996). Moreover, definitive predictions of neutron star
kick velocities from aspherical supernovae will require
three-dimensional simulations. (The aforementioned simulations were
carried out in two dimensions.) Recent simulations of
neutrino-driven convection in two and three dimensions demonstrate that,
as expected, longer-wavelength modes break up in three
dimensions, rendering the angle-averaged flow qualitatively much closer to
spherically symmetric (Knerr et al. 1998; see also M\"{u}ller 1993 and
Janka and M\"{u}ller 1996). In light of this, it is difficult to see how
the already low values for neutron star kick velocities obtained by
invoking convection and/or convection-induced anisotropic neutrino
emission during the explosion itself could be enhanced when these same
simulations are carried out in three dimensions.  

In this paper, we compare three-flavor multigroup Boltzmann
neutrino transport ({\small MGBT}) and (Bruenn's) {\small MGFLD} in 
spherically symmetric, hydrostatic, thermally frozen, postbounce
profiles, with an eye toward quantities central to the postbounce 
neutrino heating mechanism for reviving the stalled shock. In 
particular, for both transport methods, we compute and compare 
the neutrino luminosities, {\small RMS} energies, mean inverse flux 
factors, and net heating rates as functions of radius, time, and precollapse model.
We then discuss the ramifications our results have for the supernova 
mechanism. This work is a continuation and extension of core collapse simulations
(Mezzacappa and Matzner 1989, Mezzacappa and Bruenn 1993a,b,c), in 
which exact Boltzmann neutrino transport and multigroup flux-limited 
diffusion were compared.

\section{Initial Models, Codes, and Methodology}

We begin with 15 ${\rm M}_{\odot}$ and 25 ${\rm M}_{\odot}$ precollapse models
 S15s7b and S25s7b 
provided by Woosley (1995). The initial models were evolved through 
core collapse and bounce using one-dimensional Lagrangian hydrodynamics and 
{\small MGFLD} neutrino transport  
coupled to  the Lattimer--Swesty equation of
state (Lattimer \& Swesty 1991). The data at 106 ms 
 and
233 ms after bounce for S15s7b and 156 ms after bounce for S25s7b were thermally 
 and hydrodynamically frozen. 
Stationary-state neutrino distributions were computed for
these profiles using both {\small MGBT} and 
{\small MGFLD}. 

The {\small MGBT} simulations were performed using
{\small BOLTZTRAN}: a Newtonian gravity,  $O(v/c)$, three-flavor, Boltzmann neutrino transport
code developed for the supernova problem and used thus far for studies of stellar 
core collapse (Mezzacappa \& Matzner 1989, Mezzacappa \& Bruenn 1993abc). The {\small
MGFLD} simulations were performed using {\small MGFLD-TRANS}: a Newtonian gravity,  $O(v/c)$,
 three-flavor, {\small MGFLD} neutrino transport code, which
has been used for both core collapse and postbounce evolution 
(e.g., Bruenn 1985, 1993). 

The {\small MGBT} simulations used 110 nonuniform spatial 
zones spanning radii from the origin to 4744 km and 4673 km for model S15s7b at 
 $t_{\rm pb}=106$ ms and 233 ms respectively, and to 
2096 km for model S25s7b at $t_{\rm pb}=156$ ms.  
 Twelve energy 
zones spanning a range between 5 and 225 MeV were used to resolve the neutrino spectra. The 
{\small MGFLD} used the same spatial and energy grids.  
Simulations with 
20 energy zones spanning the same energy range were performed with 
{\small BOLTZTRAN}; no changes in the results presented here were seen.

For the {\small MGBT} simulations there is an added dimension:  neutrino direction 
cosine.  Because {\small MGBT} computes the neutrino distributions 
as a function of direction cosine and energy for each spatial zone, the isotropy 
of the neutrino radiation field as 
a function of radius and neutrino energy is computed from first principles.
This is one of the key features distinguishing {\small MGBT} and {\small MGFLD}.  Because 
the isotropy of the neutrino radiation field is critical to the shock reheating/revival, four Gaussian 
quadrature sets (2--, 4--, 6--, and 8--point) were implemented in the {\small MGBT} simulations to ensure 
numerical convergence of the results. 

\section{\bf Results}

For electron neutrino and antineutrino absorption on neutrons and protons, 
the neutrino heating rate (in MeV/nucleon) in the region between the neutrinospheres 
and the shock can be written as

\begin{equation}
\dot{\epsilon}=\frac{X_{n}}{\lambda_{0}^{a}}\frac{L_{\nu_{\rm e}}}{4\pi r^{2}}
                <E^{2}_{\nu_{\rm e}}><\frac{1}{\sf F}>
              +\frac{X_{p}}{\bar{\lambda}_{0}^{a}}\frac{L_{\bar{\nu}_{\rm e}}}{4\pi r^{2}}
                <E^{2}_{\bar{\nu}_{\rm e}}><\frac{1}{\bar{\sf F}}>,
\label{eq:heatrate}
\end{equation} 
\smallskip

\noindent where: $X_{n,p}$ are the neutron and proton fractions; 
$\lambda_{0}^{a}=
\bar{\lambda}_{0}^{a}= {G_{F}^2}{\rho}{(g_{V}^2+3g_{A}^2)}/{\pi}{(hc)^4}{m_{B}}$; 
 $G_{F}/(\hbar c)^{3}=1.166\times10^{-5}$ GeV$^{-2}$ is the Fermi 
coupling constant; $\rho$ is the 
matter density; 
$g_{V}=1.0$, $g_{A}=1.23$; $m_{B}$ is the baryon mass;  
and $L_{\nu_{\rm e},\bar{\nu}_{\rm e}}$, $<E^{2}_{\nu_{\rm e},\bar{\nu}_{\rm e}}>$, 
and ${\sf F},\bar{\sf F}$ are the electron neutrino and antineutrino luminosities, {\small RMS} 
energies, and mean inverse flux factors, defined by

\begin{equation}
L_{\nu_{\rm e}}=4\pi r^{2}\frac{2\pi c}
                               {(hc)^{3}}\int dE_{\nu_{\rm e}} d\mu_{\nu_{\rm e}} E^{3}_{\nu_{\rm e}} \mu_{\nu_{\rm e}} f, 
\label{eq:nuelumin}
\end{equation}

\begin{equation}
\langle E^{2}_{\nu_{\rm e}}\rangle = \frac{\int dE_{\nu_{\rm e}}d\mu_{\nu_{\rm e}} E^{5}_{\nu_{\rm e}} f}
                                          {\int dE_{\nu_{\rm e}}d\mu_{\nu_{\rm e}} E^{3}_{\nu_{\rm e}} f}, 
\label{eq:nuerms}
\end{equation}

\begin{equation}
\langle \frac{1}{\sf F}\rangle =\frac{\int dE_{\nu_{\rm e}} d\mu_{\nu_{\rm e}} E^{3}_{\nu_{\rm e}} f}
                                     {\int dE_{\nu_{\rm e}} d\mu_{\nu_{\rm e}} E^{3}_{\nu_{\rm e}} \mu_{\nu_{\rm e}} f}
=\frac{cU_{\nu_{\rm e}}}{F_{\nu_{\rm e}}}. 
\label{eq:nuefluxfac}
\end{equation}
\smallskip

\noindent In equations (\ref{eq:nuelumin})--(\ref{eq:nuefluxfac}), 
$f$ is the electron neutrino distribution function, which is a function
of the electron neutrino direction cosine, $\mu_{\nu_{\rm e}}$, and energy, $E_{\nu_{\rm e}}$.
In equation (\ref{eq:nuefluxfac}), $U_{\nu_{\rm e}}$ and $F_{\nu_{\rm e}}$
are the electron neutrino energy density and flux. Corresponding quantities 
can be defined for the electron antineutrinos. 
Success or failure to generate explosions via neutrino reheating must ultimately rest on the 
three quantities defined in equations (\ref{eq:nuelumin})--(\ref{eq:nuefluxfac}). 
Both the {\small MGBT} and the {\small MGFLD} stationary state distributions were computed
in the same thermally and hydrodynamically frozen matter configuration. 

In Figure 1, at 233 ms after bounce for model S15s7b, 
 we plot the electron neutrino and antineutrino {\small RMS} energies, luminosities, and mean 
inverse flux factors as functions of radius for our 
(8-point Gaussian quadrature) {\small MGBT} 
 and {\small MGFLD}
runs.
Energy-averaged electron neutrino- and 
antineutrino-spheres were located by calculating an energy-integrated neutrino  
depth defined by

\begin{equation}
\overline{\tau}=\frac{\int^{r}_{\infty}dr^{'}\int^{\infty}_{0}dE_{\nu}
d\mu_{\nu}E_{\nu}^{3}f/{\lambda}}{\int^{\infty}_{0}dE_{\nu}d\mu_{\nu}E_{\nu}^{3}f},
\label{eq:taubar}
\end{equation}
\smallskip

\noindent and determining the radius at which $\overline{\tau}=2/3$.  
The neutrinospheres (at 57 km and 48 km, for electron neutrinos and antineutrinos, respectively),
 and the location of the shock (at 191 km), are indicated by arrows. 
The gain radius (neutrino-energy integrated), located at 98 km, is also marked by an arrow.
For the electron neutrinos, the differences in {\small RMS} 
energies between {\small MGBT} and {\small MGFLD} are at most 
2\% throughout most of the region plotted, although  
{\small MGBT} consistently gives higher energies.  The 
differences between {\small MGBT} and {\small MGFLD} antineutrino {\small RMS} energies
are smaller, and neither transport scheme yields consistently higher values.  
It should be noted that we expect larger
differences when a fully hydrodynamic simulation is carried
out, with {\small MGBT} giving harder spectra (Mezzacappa and Bruenn 1993a,c;
 see also Burrows 1998). In
a static matter configuration,
differences that result from different treatments of the neutrino energy
shift measured by comoving observers do not occur.

Significant differences between {\small MGBT} and {\small MGFLD} are evident when comparing 
the neutrino and antineutrino luminosities and mean inverse flux factors.  
The luminosity curves for both electron neutrinos and antineutrinos coincide for both transport
methods until the neutrinospheres are approached from below.  Just below the neutrinospheres, 
the {\small MGBT} luminosities diverge upward from the {\small MGFLD} luminosities, 
differing by 7\% (4\% for antineutrinos) at the neutrinospheres.  
The root cause of this difference is that 
the {\small MGFLD} interpolation underestimates the flux in this region. 
After a decline from this 
maximum difference, the fractional 
difference grows from approximately 
3\% at the base of the gain region to
a constant difference of 6\% beyond about 170 km.  Similar behavior is 
exhibited by the antineutrino luminosities, 
with the same fractional differences, 3\% and 6\%,  seen at the base of the gain region and near the shock, 
respectively.  

For the electron neutrinos, the fractional difference between $<1/{\sf F}>_{\rm \tiny MGFLD}$ and
$<1/{\sf F}>_{\rm \tiny MGBT}$
  is 2\%, 8\%, and 12\% at the neutrinosphere, gain radius, and shock, respectively.
Just above the shock, the difference converges to 10\%,
 and is maintained to the edge of the core.
Focusing on the semitransparent region, 
 $<1/{\sf F}>_{\rm \tiny MGFLD}$ is greater  
until the gain radius is approached from below; i.e., the 
{\small MGFLD} neutrino radiation field is more isotropic 
than the {\small MGBT} radiation field below these radii.  
At 80 km, as the gain radius is approached, {\small MGFLD}
computes a sharp decrease in $<1/{\sf F}>$. 
Looking at    
 Figure 2, where we plot
the density and the sum of the {\small MGBT} electron neutrino and antineutrino luminosities
for this time slice
 as functions of radius, it is evident as the gain radius is approached from below that 
the luminosity sum begins to turn over, marking the 
enclosure of the neutrino and antineutrino source.  
Therefore, for {\small MGFLD}, the accelerated transition to free
streaming occurs not at the neutrinospheres, as might have been anticipated,
but at a radius within which most of the neutrino and antineutrino source is enclosed.
For example, for this time slice the electron neutrino luminosity at the neutrinosphere 
is only 76\%  of its peak value.

In Figure 1, there is a second dip and a small spike in $<1/{\sf F}>_{\rm \tiny MGFLD}$ at 106 km 
 and 163 km, respectively; $<1/{\sf F}>_{\rm \tiny MGBT}$ is smooth through these radii.  
Again, examining Figure 2, 
the density shows discontinuities at 106 km and 
 163 km,  
which produce these  
features.  The density profile flattens 
at  106 km  and then drops precipitously.  
There is a corresponding flattening and sharp drop in $<1/{\sf F}>_{\rm \tiny MGFLD}$ in this region.
 The density actually increases at 160 km, then immediately falls off. This results in an increase in 
$<1/{\sf F}>_{\rm \tiny MGFLD}$ at that radius, followed by a sharp decrease. 
  In both cases, the isotropy of the {\small MGFLD} 
neutrino radiation field is altered by local conditions.   

For the electron antineutrinos, the same features are seen in $<1/{\sf F}>_{\rm \tiny MGFLD}$. 
  The fractional difference  
is 0\%, 11\%, and 11\% at the neutrinosphere, gain radius, and shock, respectively.
 The initial sharp decrease in 
$<1/{\sf F}>_{\rm \tiny MGFLD}$ occurs at a smaller radius.    
 The antineutrino luminosity maximum, i.e., the point at which the antineutrino source is enclosed, 
is at a smaller radius.

In Figure 3, we plot the same three quantities for the earlier time slice in 
our 15 M$_\odot$ model ($t_{\rm pb}=106$ ms).  
The differences between {\small MGBT} and {\small MGFLD} 
are similar to those seen at the later postbounce
time ($t_{\rm pb}=233$ ms).  
The electron neutrino {\small RMS} energies are slightly higher for {\small MGBT}, again  
by about 2\%. 
 The differences in antineutrino {\small RMS} energies are again variable in both sign and magnitude, 
but are never more than 2\%.   
  The difference in luminosity is 11\% (8\% for antineutrinos) at the neutrinosphere, 
 11\% (7\% for antineutrinos) at the gain radius,
  and settles to a constant difference of 9\% (6\% for
antineutrinos) above 170 km.  As in the later time slice, {\small MGFLD} underestimates the 
flux beginning below the neutrinospheres, extending 
everywhere above the neutrinospheres, which in turn results in a lower luminosity.

For
electron neutrinos, the fractional
difference between $<1/{\sf F}>_{\rm \tiny MGFLD}$ and $<1/{\sf F}>_{\rm \tiny MGBT}$ 
is 4\%, 2\%,  and 17\%  
at the neutrinosphere, gain radius, and shock, respectively.  
Most important, the same abrupt
decrease in $<1/{\sf F}>_{\rm \tiny MGFLD}$ beginning just above the gain radius is evident.    
In Figure 4, we plot the density and the sum of the {\small MGBT} 
electron neutrino and antineutrino luminosities for $t_{\rm pb}=106$ ms.  The decrease in  
$<1/{\sf F}>_{\rm \tiny MGFLD}$ again occurs near the radius where the luminosity sum 
turns over:  128 km.     
There is also a small dip in  $<1/{\sf F}>_{\rm \tiny MGFLD}$ 
  at 172 km.  
 Looking at Figure 4, the only significant dip in density occurs at 172 km,  
 causing a local 
decrease in the isotropy of the {\small MGFLD} radiation field. 

For the electron antineutrinos, the difference between  $<1/{\sf F}>_{\rm \tiny MGFLD}$ and
$<1/{\sf F}>_{\rm \tiny MGBT}$  
is  3\%, 2\%, and 16\% at the neutrinosphere, gain radius, and shock, respectively. 
The point at which $<1/{\sf F}>_{\rm \tiny MGFLD}$ drops below $<1/{\sf F}>_{\rm \tiny MGBT}$ is
translated inwards, as expected, towards the antineutrino luminosity maximum.
For both electron neutrinos and antineutrinos, 
 a difference of about 13\% is maintained 
with increasing radius above the shock. 
  
  Figure 5 contains the same information as Figures 1 and 3, but for our 25 M$_\odot$ model
at 156 ms after bounce. 
 The similarities between the  
 25 M$_\odot$ and the 15 M$_\odot$ results are striking,  
considering the marked difference in core structure.  
 The identical trend in {\small RMS} energies is again seen:  
there are small differences for the electron neutrinos, but consistently higher values
  are obtained with {\small MGBT} (again, $\sim$ 2\% higher);  
there are smaller, sometimes oscillating, differences between {\small MGBT} and {\small MGFLD} 
for the electron antineutrinos.  
The absolute value of the neutrino luminosities at the gain radius is 
almost a factor of 2 greater than in either of the 15 M$_\odot$ slices, yet 
the relative differences between the luminosities computed 
by the two transport methods are similar.
The electron neutrino luminosities differ by 9\% (5\% for antineutrinos) at the 
neutrinosphere,
 11\% (9\% for antineutrinos) at the gain radius, and  
 9\% (7\% for antineutrinos) at the shock.  The difference at the shock 
is maintained to the edge of the core. Again, this difference in luminosity 
is caused by an underestimation of the flux by {\small MGFLD}.  

Among the three slices considered, the 25 M$_\odot$ core
 gives rise to the most dramatic differences in $<1/{\sf F}>$.  
 For electron neutrinos, we find fractional differences of 1\%, 2\%, and 24\% in $<1/{\sf F}>$           
 at the neutrinosphere, gain radius, and shock, respectively. 
There is a sharp decrease in $<1/{\sf F}>_{\rm \tiny MGFLD}$ at 109 km.  As in the 15 M$_\odot$ case, 
this change in $<1/{\sf F}>_{\rm \tiny MGFLD}$ is correlated with 
the enclosure of the neutrino source. 
This correlation is evident in Figure 6, where we plot the 
density and sum of the {\small MGBT} electron neutrino and antineutrino luminosities for
this time slice. Note that the {\small MGBT} luminosity sum begins to turn over 
near 109 km, coincident with the sharp decrease in $<1/{\sf F}>_{\rm \tiny MGFLD}$. 

There are other precipitous drops in $<1/{\sf F}>_{\rm \tiny MGFLD}$ 
at 125 km and 162 km.  
Also apparent in Figure 6  are
precipitous drops in density at these radii.  These drops in density produce changes 
in the local neutrino radiation field computed by {\small MGFLD}.    

For electron antineutrinos, $<1/{\sf F}>_{\rm \tiny MGFLD}$ 
exhibits similar structure, with fractional differences of 
0\%, 1\%, and 19\% at the neutrinosphere, gain radius, and shock, respectively. 
$<1/{\sf F}>_{\rm \tiny MGFLD}$ for antineutrinos also contains three dips:  
$<1/{\sf F}>_{\rm \tiny MGFLD}$ drops below $<1/{\sf F}>_{\rm \tiny MGBT}$ at 105 km, 
where most of the antineutrino
source is enclosed, and additional 
dips at 125 km and 162 km are also evident, again resulting from the density dips 
visible in Figure 6.  
 For both electron neutrinos and antineutrinos,  constant 
differences $\sim$ 12\% are seen above the shock.

Because each of the quantities plotted in Figures 1, 3, and 5 is consistently greater for 
{\small MGBT} (while this is not strictly true  
for the antineutrino {\small RMS} energies in our stationary state comparisons, in a fully 
dynamical simulation these energies will be consistently higher for {\small MGBT}
[Mezzacappa and Bruenn 1993a,c;
 see also Burrows 1998]), 
and because the neutrino 
heating rate is proportional to each of them, 
 {\small MGBT} yields a significantly higher heating rate.  As an example, just above the
gain radius for model S15s7b at $t_{\rm pb}=233$ ms, at the point of peak heating, 
{\small MGBT} yields a heating rate from neutrino absorption that is $(102\%)^{2} \times 110\% \times 112\%$
 of the {\small MGFLD} rate. 
Note in equation (\ref{eq:heatrate}) that the heating rate depends linearly on {\it both} 
the  
neutrino luminosity and the mean inverse flux factor. 
  A reliable determination of the
heating rate in and around the gain region therefore depends on a realistic solution of the transport equation
 in which both of these quantities are determined self consistently.

In 
Figure 7, for   
{\small MGBT} and {\small MGFLD}, 
we plot the {\it net} neutrino heating rates as functions of radius 
for model S15s7b at $t_{\rm pb}=233$ ms. 
(As described in Section 2, the results from four Gaussian quadrature sets are plotted to demonstrate numerical
convergence.) 
These rates include the
contributions from both the electron neutrinos and antineutrinos, and were computed using 
the following formulae: 

\begin{equation}
(d\epsilon /dt)_{i}=c\int E_{\nu}^{3}dE_{\nu}
[\psi^{0}_{i}/\lambda^{(a)}_{i}-j_{i}(1-\psi^{0}_{i})]/\rho (hc)^{3} 
\end{equation}

\noindent where $\epsilon$ is the internal energy per gram; 
$E_{\nu}$, $\psi^{0}_{i}$, $\lambda^{(a)}_{i}$, and $j_{i}$ 
are the electron neutrino or antineutrino energy, zeroth angular 
moment, absorption mean free path, 
and emissivity, respectively; 
$i=1$ corresponds to electron neutrinos, and 
$i=2$ corresponds to electron antineutrinos. 
Only the contributions from neutrino emission and absorption
were included: in our models, contributions from neutrino--electron scattering 
and other processes contribute only 
a few percent to the neutrino heating rate at and before  
233 ms after bounce; they become more important ($\sim$15\%) 
at later times.
The (8--point Gaussian quadrature) {\small MGBT} simulation yields a net heating rate just above the gain radius
 that is 
$\sim$1.3 times the {\small MGFLD} rate, and a net cooling rate
below the gain radius that is consistently $\sim$0.8 times the {\small MGFLD} rate. 
Comparing the net heating rate to the luminosity sum in Figure 2, 
the extent of the gain region (from 98 km to 180 km ) is well correlated with 
the region between the luminosity maximum and the radius where 
the luminosity levels off
 to a constant value (from 101 km to 170 km).  
Note also that the gain radius is located at a smaller radius for {\small MGBT}, and  
consequently, the size of the net heating region below the shock is somewhat larger. 
 
For (8--point Gaussian quadrature) {\small MGBT} and {\small MGFLD}, 
Figure 8 shows the net heating curves for S15s7b at an earlier postbounce time, $t_{\rm pb}=106$ ms. 
 The region between the neutrinospheres and 
the shock is a bit smaller:  The shock is approximately at the same radius, but 
the energy-averaged electron neutrinosphere is 
closer to the shock (at 84 km, versus 58 km for $t_{\rm pb}=233$ ms).
Similar net heating profiles are seen, but the 
differences between {\small MGBT} and {\small MGFLD} are even greater. {\small MGBT} yields 
a net heating rate that 
is $\sim$2 times the {\small MGFLD} rate just above the base of the gain region.
  Again, the cooling rate is consistently $\sim$0.8 times the {\small MGFLD} rate below 
the gain radius.  
The correlation between the gain region and the region between the luminosity-sum maximum 
and the point at which the luminosity sum is constant is also quite strong for this time slice. 
 The gain radius (103 km)  is just below 
the luminosity maximum (107 km);  also, the luminosity levels off at a radius (173 km)  
near  the upper edge of the gain region (182 km).   


Figure 9 contains the net heating curves for our 
 25 M$_{\odot}$ model at $t_{\rm pb}=156$ ms.  Again, in this case the 
(8--point Gaussian quadrature) {\small MGBT} net heating  
rate is  $\sim$2 times the {\small MGFLD} rate just above the base of the gain region, 
and the net cooling rate below the gain 
radius is consistently $\sim$0.8 times the {\small MGFLD} rate.  
The gain radius (111 km) and the luminosity-sum maximum (108 km) are strongly
correlated, as are the radius marking the upper extent of the gain 
region (180 km) and the radius at which the luminosity sum levels off (170 km). 
 Although the differences in the net heating rate are comparable to those seen in 
the $t_{\rm pb}=106$ ms slice for model S15s7b, they 
occur at a significantly later postbounce time. 
At a given postbounce time, {\small MGBT} 
 provides the greatest enhancement in 
net heating for larger 
progenitor masses.

\section{\bf Summary, Discussion, and Conclusions}

Comparing three-flavor {\small MGBT} and three-flavor 
{\small MGFLD} in postbounce supernova environments, we find that 
{\small MGBT} leads to a significant increase/decrease in the 
 {\it net} heating/cooling rate, particularly above/below 
 the gain radius. The {\small MGBT} net heating  rate can be 
as much as $\sim$2 times the {\small MGFLD} net heating rate above the gain radius, with
net cooling rates that are typically $\sim$0.8 times the {\small MGFLD} rate below 
the gain radius.  These differences 
 stem primarily from differences in the neutrino luminosities and mean inverse flux 
factors;  the heating rate 
is linearly proportional to both these quantities, and differences in both add to 
produce a significant difference in the net heating rate. 

In Figure 10,  we plot the sum of the electron neutrino and antineutrino luminosities 
computed in the  {\small MGFLD} S15s7b dynamic simulation for several different postbounce 
times.  It is important to note that   
 the total luminosity changes by 
$\sim$5--15\% between 100 km and 200 km on time scales $\sim$30 ms.  Moreover, the light crossing time 
between 100 km and 200 km is $\sim$1/3 ms.  Therefore, the neutrino source in our simulations changes on 
time scales that are two orders of magnitude greater than the time scales on which stationary state
is established in this region. 
This suggests our stationary state results closely reflect what will occur in dynamic simulations.

We also observe that the differences in the net heating rate are greatest at 
earlier postbounce times for a given progenitor mass, 
 and greater at any given postbounce time for greater progenitor mass. 
This is illustrated in Table 1.  
  The  increase in net heating with increased progenitor mass  
is advantageous because of the slower fall-off in the preshock accretion ram pressure.  

These results have at least two important ramifications for the
supernova mechanism: 

(1) With the dramatic increase in net heating 
above the gain radius, which is seen in all of our postbounce
slices, it may be possible 
to obtain explosions in one dimension without multidimensional effects  
such as convection; 
this will be aided by the decrease in net cooling below the gain
radius.
It should be noted that our postbounce slices come from full radiation 
hydrodynamics simulations implementing {\small MGFLD} that marginally failed
to produce explosions (Bruenn 1993). 
The marginality of Bruenn's simulations is an important motivating factor in comparing
our {\small MGBT} results  solely with his {\small MGFLD} results.
All else being equal,
 increases in net heating  of the magnitude 
documented here would most likely lead to 
explosions.  
However, simulations coupling {\small MGBT} and the core
hydrodynamics must be carried out in order
to compute any feedbacks. It remains to be seen 
whether the {\small MGBT} heating rate will remain sufficiently high to generate 
an explosion.  Also, the effects of using general relativistic gravity,  
hydrodynamics, and neutrino transport must 
be explored, especially if explosions are obtained in the Newtonian limit. For example, the 
 redshifted neutrino energies, the smaller gain region, and the greater infall velocities
in the gain region will most likely 
 conspire to make explosions more difficult to obtain (Mezzacappa et al. 1998b, 
DeNisco, Bruenn, \& Mezzacappa 1998, Bruenn, DeNisco, \& Mezzacappa 1998).   
 
(2) With the dramatic increase in net heating 
occurring near the base of the gain region,  
 we anticipate that {\small MGBT} 
 coupled to two-dimensional hydrodynamics will yield 
more vigorous neutrino-driven convection than was seen in Mezzacappa
et al. (1998b), where two-dimensional 
hydrodynamics was coupled to one-dimensional {\small MGFLD}. 
The combination of increased net heating and more vigorous neutrino-driven 
convection would be more favorable for shock revival. 

In closing, our results are promising, and their ramifications 
for core collapse supernovae and, in particular, for the postbounce 
neutrino-heating, shock-revival mechanism await one- and two-dimensional 
dynamical simulations with {\small MGBT} coupled 
to the core hydrodynamics. One-dimensional simulations are currently 
underway, and we plan to report on them soon.

\section{Acknowledgements}

BM
was supported at the University of Tennessee under NASA grant 
NRA-96-04-GSFC-073. AM and MWG were supported at the Oak Ridge National Laboratory,
which is managed by Lockheed Martin Energy Research Corporation
under DOE contract DE-AC05-96OR22464, and at the University of
Tennessee, under DOE contract DE-FG05-93ER40770.
SWB was supported at Florida Atlantic University 
under NASA grant NRA-96-04-GSFC-073 and NSF grant AST-9618423. 
The simulations presented in this paper were carried out on
the Cray C90 at the National Energy Research Supercomputer
Center, the Cray Y/MP at the North Carolina Supercomputer 
Center, and the Cray Y/MP and Silicon Graphics Power
Challenge at the Florida Supercomputer Center. AM and SWB 
gratefully acknowledge the hospitality of the Institute
for Theoretical Physics, Santa Barbara, 
which is supported in part by
the National Science Foundation under grant 
PHY94-07194.  The authors also want to thank the referee, John Hayes, for 
suggestions that improved the content of this paper.

\newpage

\section{\bf References}
 
\noindent Bethe, H. \& Wilson, J. R. 1985, ApJ, 295, 14

\noindent Bowers, R. L. \& Wilson, J. R. 1982, ApJS, 50, 115

\noindent Brown, G. E., Bruenn, S. W., \& Wheeler, J. C. 1992, Comments Astrophys. 16, 153

\noindent Bruenn, S. W. 1985, ApJS, 58, 771

\noindent Bruenn, S. W. 1993, in Nuclear Physics in the Universe, eds. M. W. Guidry \& M. R. Strayer
\indent          (Bristol: IOP Publishing), p. 31

\noindent Bruenn, S. W. \& Mezzacappa, A. 1994, ApJ, 433, L45

\noindent Bruenn, S. W., DeNisco, K., \& Mezzacappa, A. 1998, ApJ, submitted

\noindent Bruenn, S. W. \& Dineva, T. 1996, ApJ, 458, L71

\noindent Burrows, A. 1998, in Proceedings of the 18th Texas Symposium on Relativistic Astrophysics,
\indent          eds. A. Olinto, J. Frieman, \& D. Schramm (Singapore: World Scientific Press), in press

\noindent Burrows, A. \& Goshy, J. 1993, ApJ, 416, L75

\noindent Burrows, A. \& Hayes, J. 1995, in High-Velocity Neutron Stars and Gamma Ray Bursts, eds.
\indent          E. Rothschild \& R. E. Lingenfelter (Woodbury, N.Y.: American Institute of Physics), p. 25

\noindent Burrows, A. \& Hayes, J. 1996, Phys.~ Rev.~ Lett.~, 76, 352

\noindent Burrows, A., Hayes, J., \& Fryxell, B. A. 1995, ApJ, 450, 830

\noindent Chevalier, R. A. 1989, ApJ, 346, 847
 
\noindent DeNisco, K., Bruenn, S. W. \& Mezzacappa, A. 1998, in Second Oak Ridge Symposium on Atomic
\indent           and Nuclear Astrophysics, ed. A. Mezzacappa 
   (Bristol: IOP Publishing), in press 

\noindent Dgani, R. \& Janka, H.-Th. 1992,  A\& A 256, 428

\noindent Herant, M. E., Benz, \& Colgate, S. A. 1992, ApJ, 395, 642

\noindent Herant, M. E., Benz, W., Hix, W. R., Fryer, C., \& Colgate, S. A. 1994, 
          ApJ, 435, 339

\noindent Janka, H.-Th. 1992, A\& A 256, 452

\noindent Janka, H.-Th., \& M\"{u}ller, E. 1994, A\& A 290, 496 

\noindent Janka, H.-Th., \& M\"{u}ller, E. 1996, A\& A 306, 167 

\noindent Keil, W., Janka, H.-Th., \& M\"{u}ller, E. 1996, ApJ, 473, L111

\noindent Knerr, J. M., Mezzacappa, A., Blondin, J. M., and Bruenn, S. W., ApJ, in preparation

\noindent Lattimer, J. M. \& Swesty , F. D. 1991, Nucl. Phys. A, 535, 331

\noindent Levermore, C. D. \& Pomraning, G. C. 1981, ApJ, 248, 321

\noindent Mezzacappa, A. \& Matzner, R. A. 1989, ApJ, 343, 853

\noindent Mezzacappa, A. \& Bruenn, S. W. 1993a, ApJ, 405, 637

\noindent Mezzacappa, A. \& Bruenn, S. W. 1993b, ApJ, 405, 669 

\noindent Mezzacappa, A. \& Bruenn, S. W. 1993c, ApJ, 410, 710

\noindent Mezzacappa, A., Calder, A. C., Bruenn, S. W., Blondin, J. M., 
          Guidry, M. W., Strayer, M. R.,\linebreak\indent \& Umar, A. S. 1998a, ApJ, 493, 848 

\noindent Mezzacappa, A., Calder, A. C., Bruenn, S. W., Blondin, J. M., 
          Guidry, M. W., Strayer, M. R.,\linebreak\indent \& Umar, A. S. 1998b, ApJ, 495, 911 

\noindent M\"{u}ller, E. 1993, in Proceedings of the 7th Workshop on Nuclear Astrophysics,
          eds. W. Hillebrandt\linebreak\indent \& E. M\"{u}ller 
   (Garching: Max-Planck-Institut f\"{u}r Astrophysik), p. 27

\noindent Miller, D. S., Wilson, J. R., \& Mayle, R. W. 1993, ApJ, 415, 278

\noindent Myra, E., Bludman, S., Hoffman, Y., Lichtenstadt, I., Sack, N., \& Van Riper, K. 1987, 
\linebreak\indent ApJ, 318, 744

\noindent Swesty, F. D. \& Lattimer, J. M. 1994, ApJ, 425, 195 

\noindent Wilson, J. R. \& Mayle, R. W. 1993, Phys. Rep., 227, 97 

\noindent Wilson, J. R. 1985, in Numerical Astrophysics, eds. J. M. Centrella, J. M. LeBlanc, 
 \linebreak\indent \& R. L. Bowers 
          (Boston: Jones \& Bartlett), p. 422 

\noindent Woosley, S. E. 1995, private communication

\newpage

\section{Figure Captions}

\figcaption{At $t_{\rm pb}=233$ ms, for model S15s7b, we plot the electron neutrino 
and antineutrino {\small RMS} energies, luminosities,
 and mean inverse flux factors versus radius for 
both Boltzmann neutrino transport and Bruenn's {\small MGFLD}.  
 The location of the energy-averaged electron neutrinosphere and anti-neutrinosphere  
 and the locations of the gain radius and shock are indicated by 
arrows.}

\figcaption{The density and {\small MGBT} electron neutrino and antineutrino luminosity  sum 
are plotted versus radius for model S15s7b at $t_{\rm pb}=233$ ms.}

\figcaption{At $t_{\rm pb}=106$ ms, for model S15s7b, we plot the electron neutrino
and antineutrino {\small RMS} energies, luminosities,
 and mean inverse flux factors versus radius for
both Boltzmann neutrino transport and Bruenn's {\small MGFLD}.
 The location of the energy-averaged electron neutrinosphere and anti-neutrinosphere 
 and the location of the gain radius and shock are indicated by                    
arrows.}
 
\figcaption{The density and {\small MGBT} electron neutrino and antineutrino luminosity  sum
are plotted versus radius for model S15s7b at $t_{\rm pb}=106$ ms.}

\figcaption{At $t_{\rm pb}=156$ ms, for model S25s7b, we plot the electron neutrino
and antineutrino {\small RMS} energies, luminosities,
 and mean inverse flux factors versus radius for
both Boltzmann neutrino transport and Bruenn's {\small MGFLD}.
 The location of the energy-averaged electron neutrinosphere and anti-neutrinosphere 
 and the location of the gain radius and the shock are indicated by                    
arrows.}

\figcaption{The density and {\small MGBT} electron neutrino and antineutrino luminosity  sum 
are plotted versus radius for model S25s7b at $t_{\rm pb}=156$ ms.}

\figcaption{At $t_{\rm pb}=233$ ms, for model S15s7b,  the net neutrino heating rates are plotted versus
radius for both Boltzmann neutrino transport and {\small MGFLD}. 
 The location of the energy-averaged electron neutrinosphere
 and the location of the shock are indicated by arrows.
The results from four Gaussian quadrature sets are plotted to demonstrate numerical convergence.}

\figcaption{At $t_{\rm pb}=106$ ms, for model S15s7b, the net neutrino heating 
rates for {\small MGFLD} and (8-point Gaussian quadrature) Boltzmann transport are plotted versus radius,   
 along with the locations of the energy-averaged electron neutrinosphere and the shock.} 

\figcaption{At $t_{\rm pb}=156$ ms, for model S25s7b, the net neutrino heating
rates for {\small MGFLD} and (8-point Gaussian quadrature) Boltzmann transport are plotted versus radius, 
along with the locations of the energy-averaged electron neutrinosphere and the shock.}

\figcaption{The sum of the electron neutrino and antineutrino luminosities from the {\small MGFLD} 
dynamic run for model S15s7b are plotted versus radius.  The luminosities from five different 
postbounce times are shown.}

\newpage
\section{Tables}
\begin{center}
\begin{table}[hb]
\footnotesize\rm
  \caption{Maximum Net Heating/Cooling Rates \label{t1}}
      \begin{tabular}{lccc}  
  \tableline
  \tableline
            {\small Progenitor Mass [${\rm M}_{\odot}$]} &
            {\small $t_{pb}$ [ms]} &
            {\small Maximum Net Heating Ratio} &
            {\small Maximum Net Cooling Ratio} 
       \nl 
      \tableline
            15&
            106&
            2.0&
            0.8
        \nl        
             &
            233&
            1.3&
            0.8
        \nl        
            25 &
            156&
            2.0&
            0.8
      \nl
   \tableline
\end{tabular}
\end{table}
\end{center}

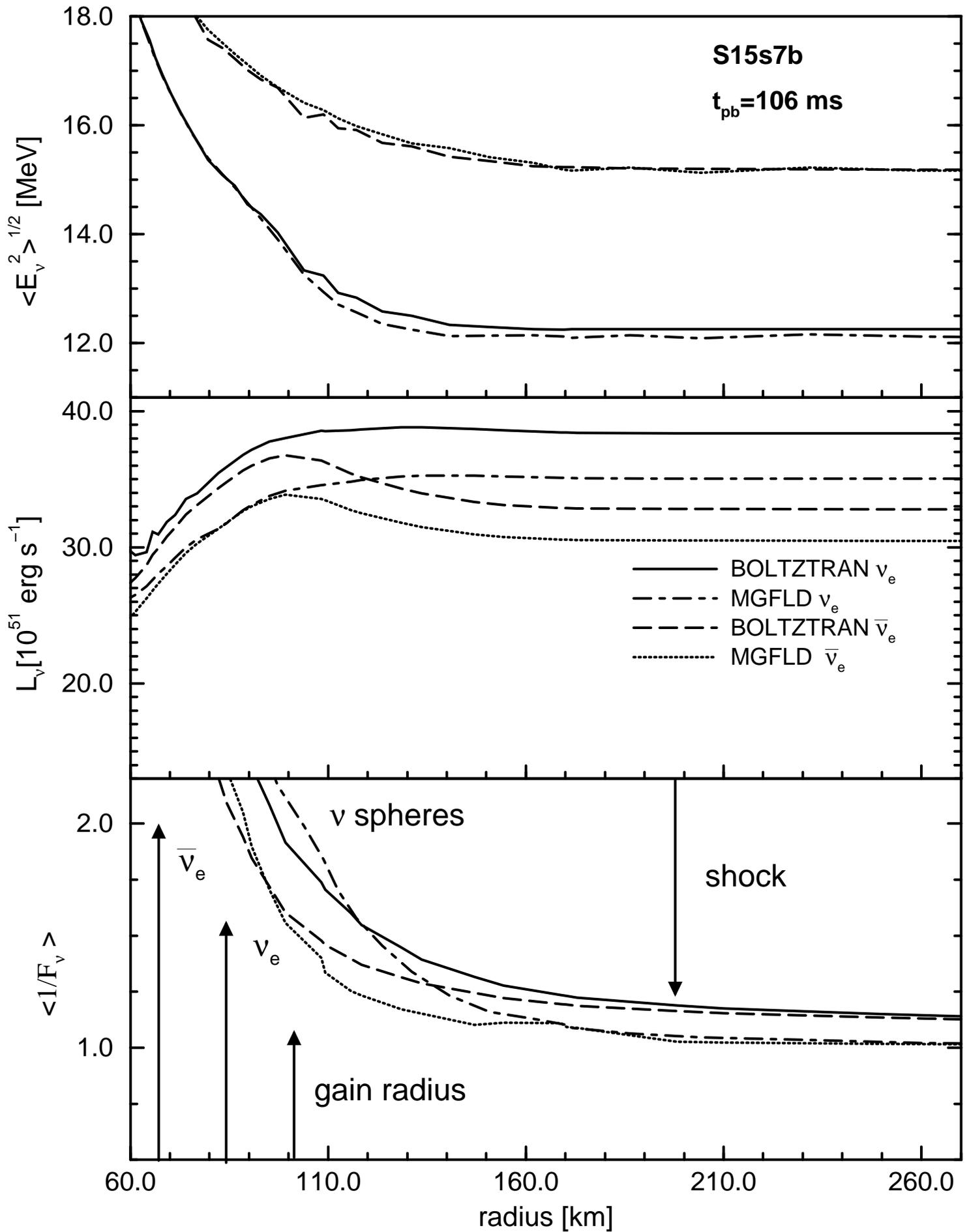
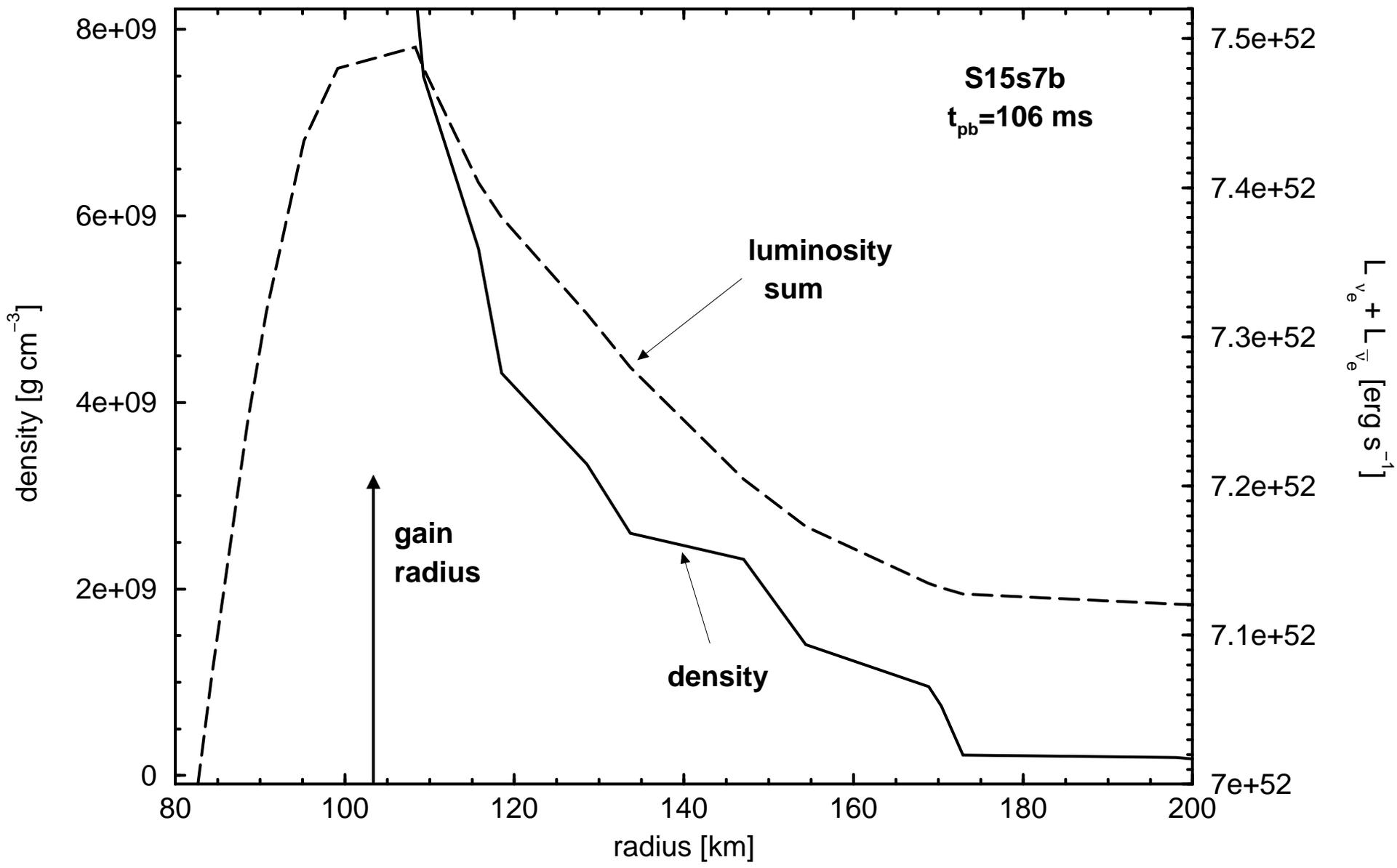
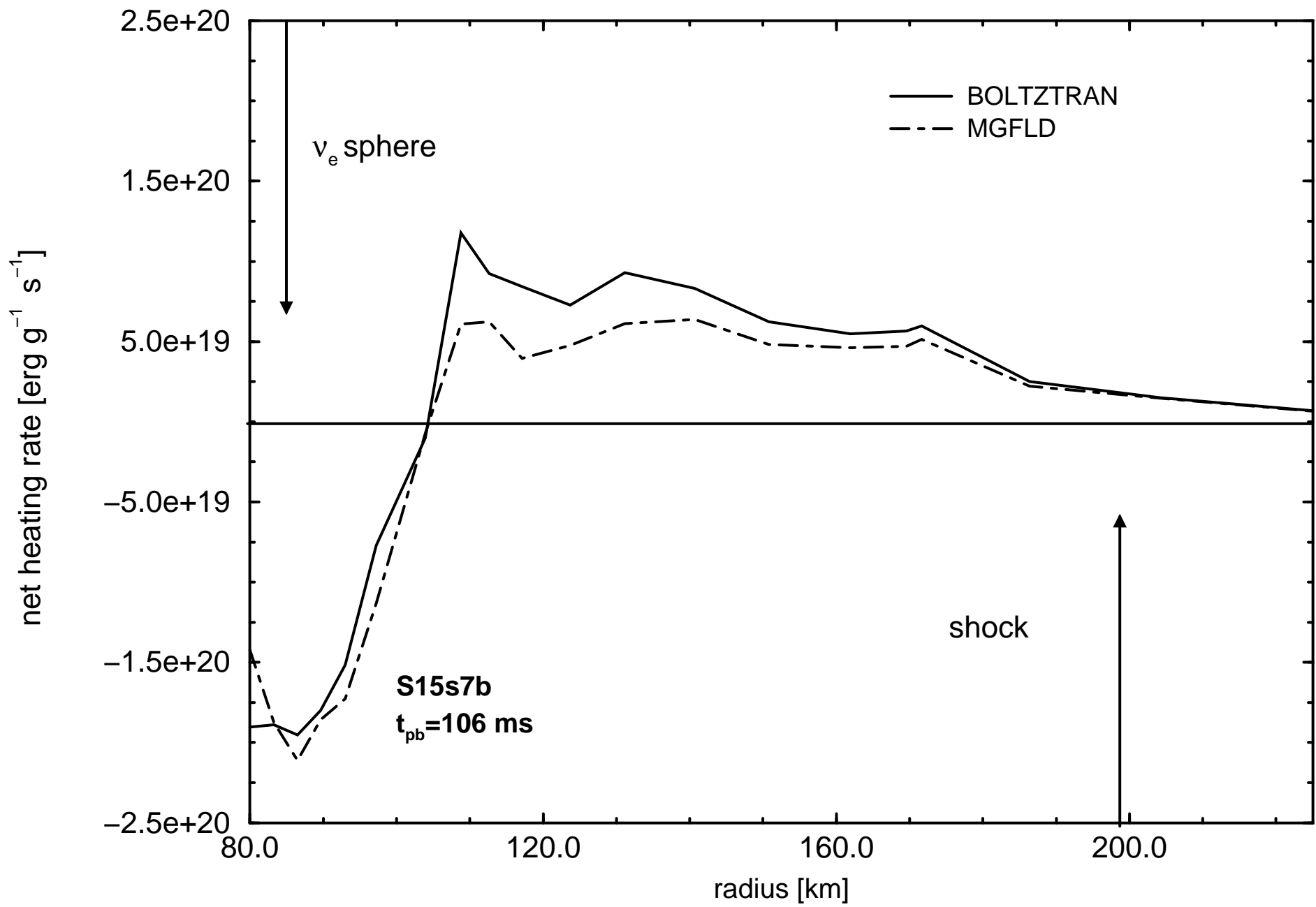
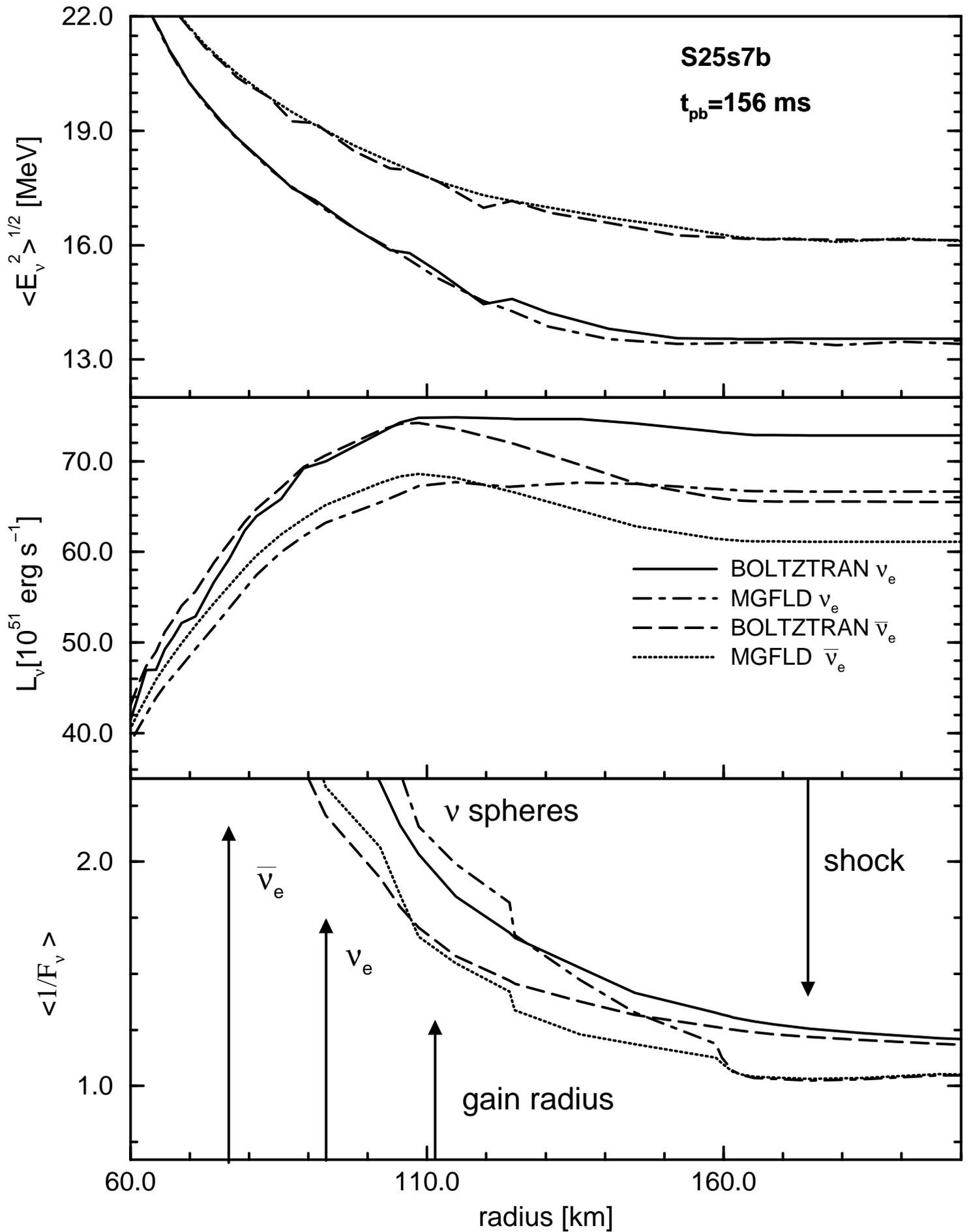
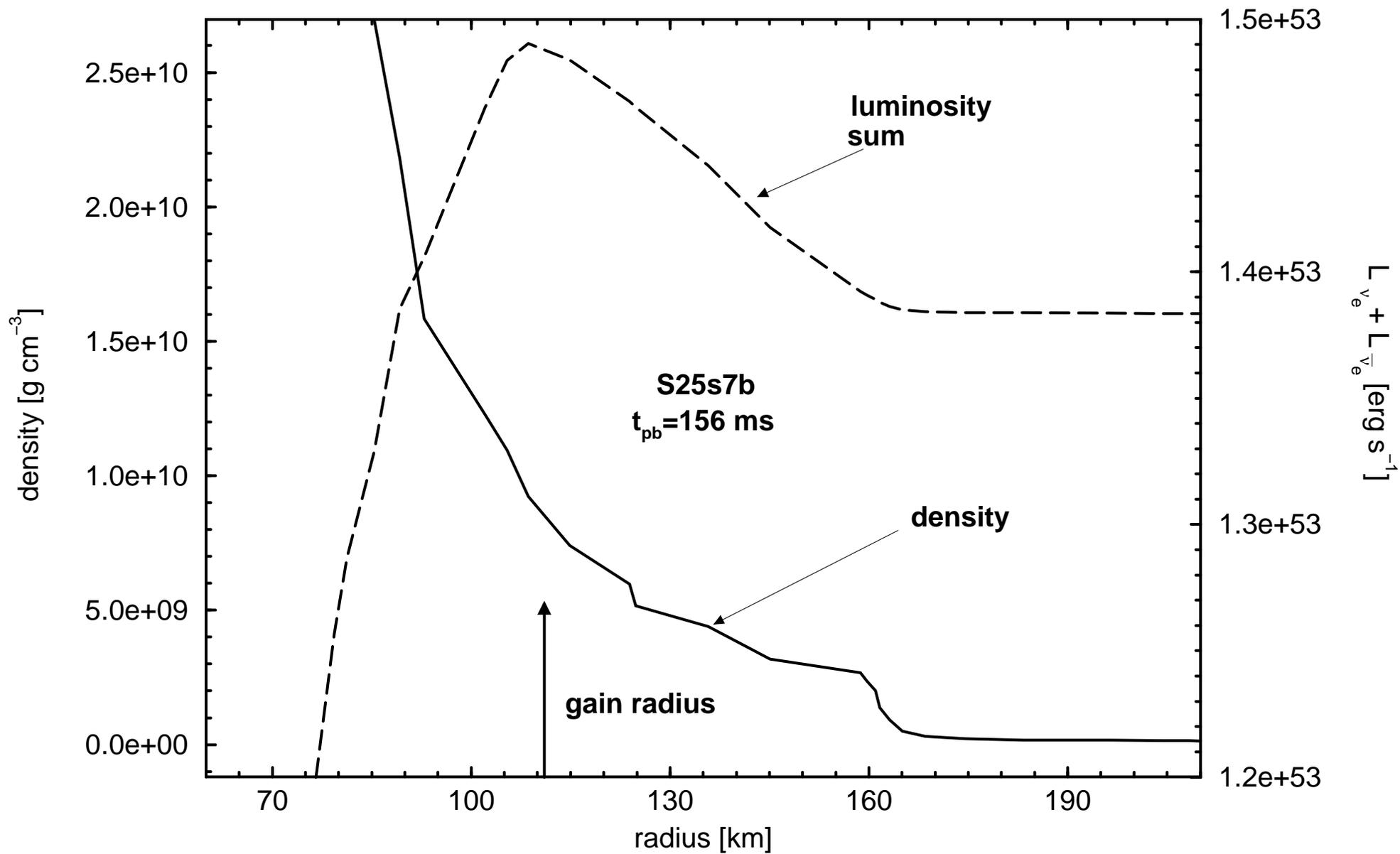
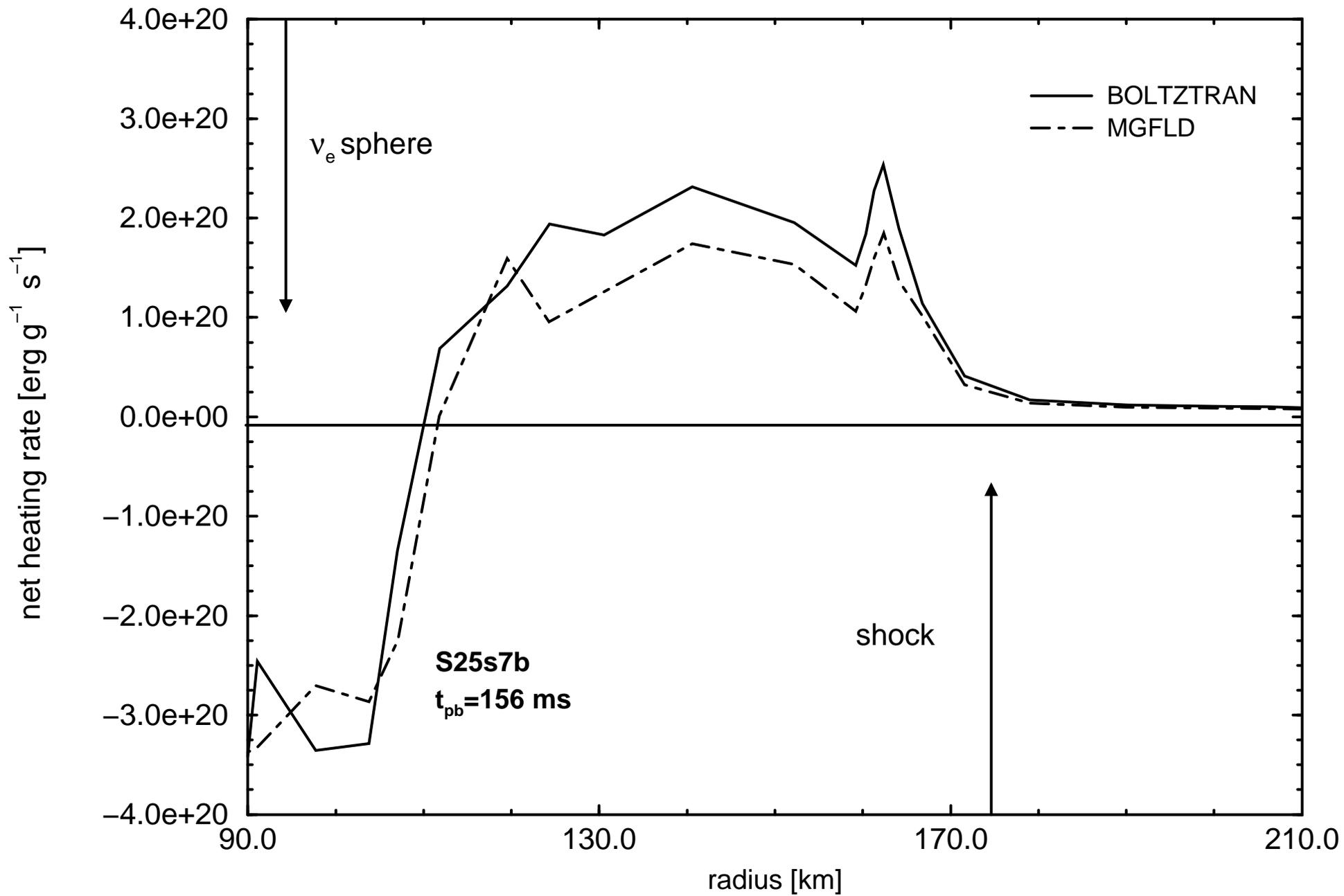
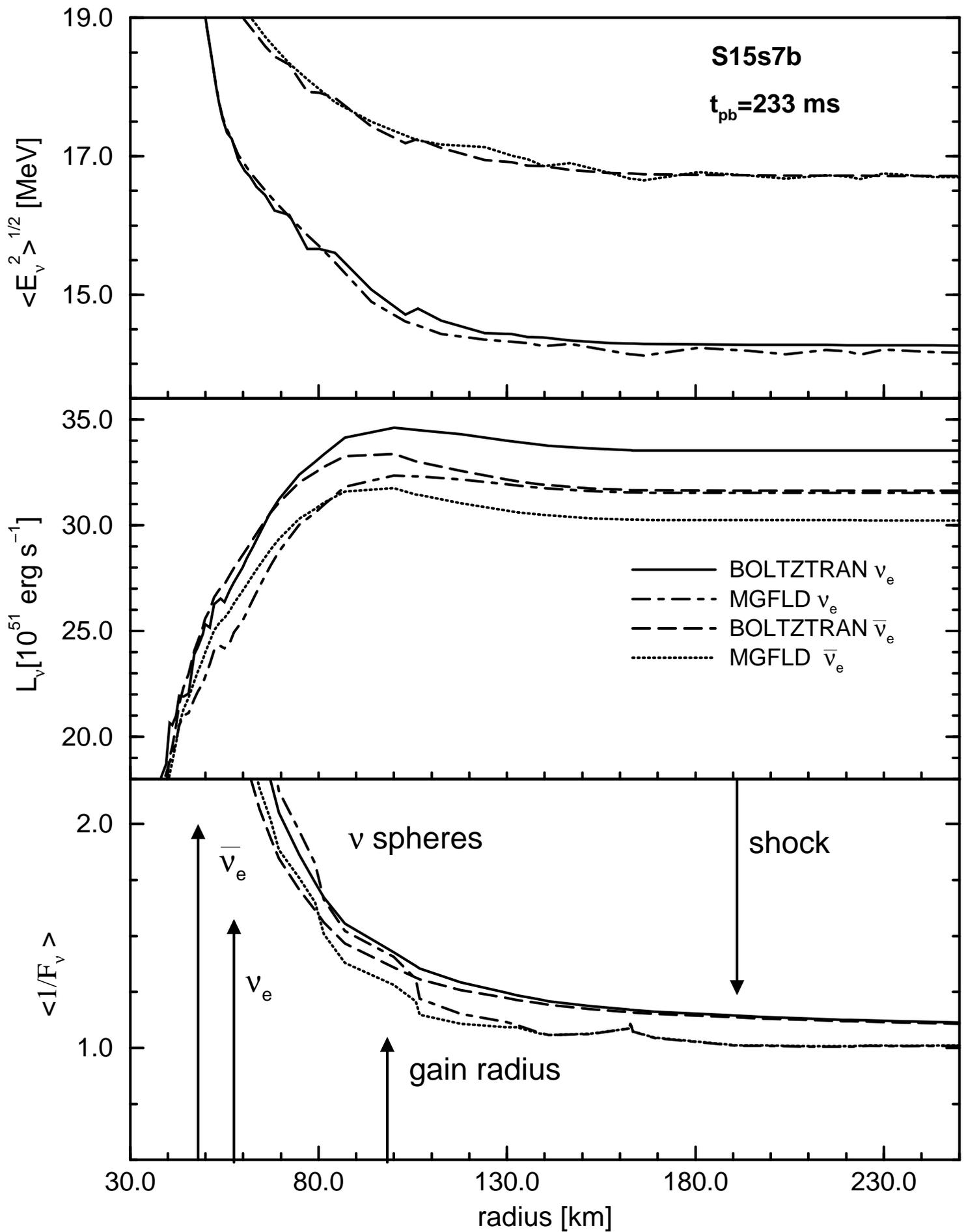
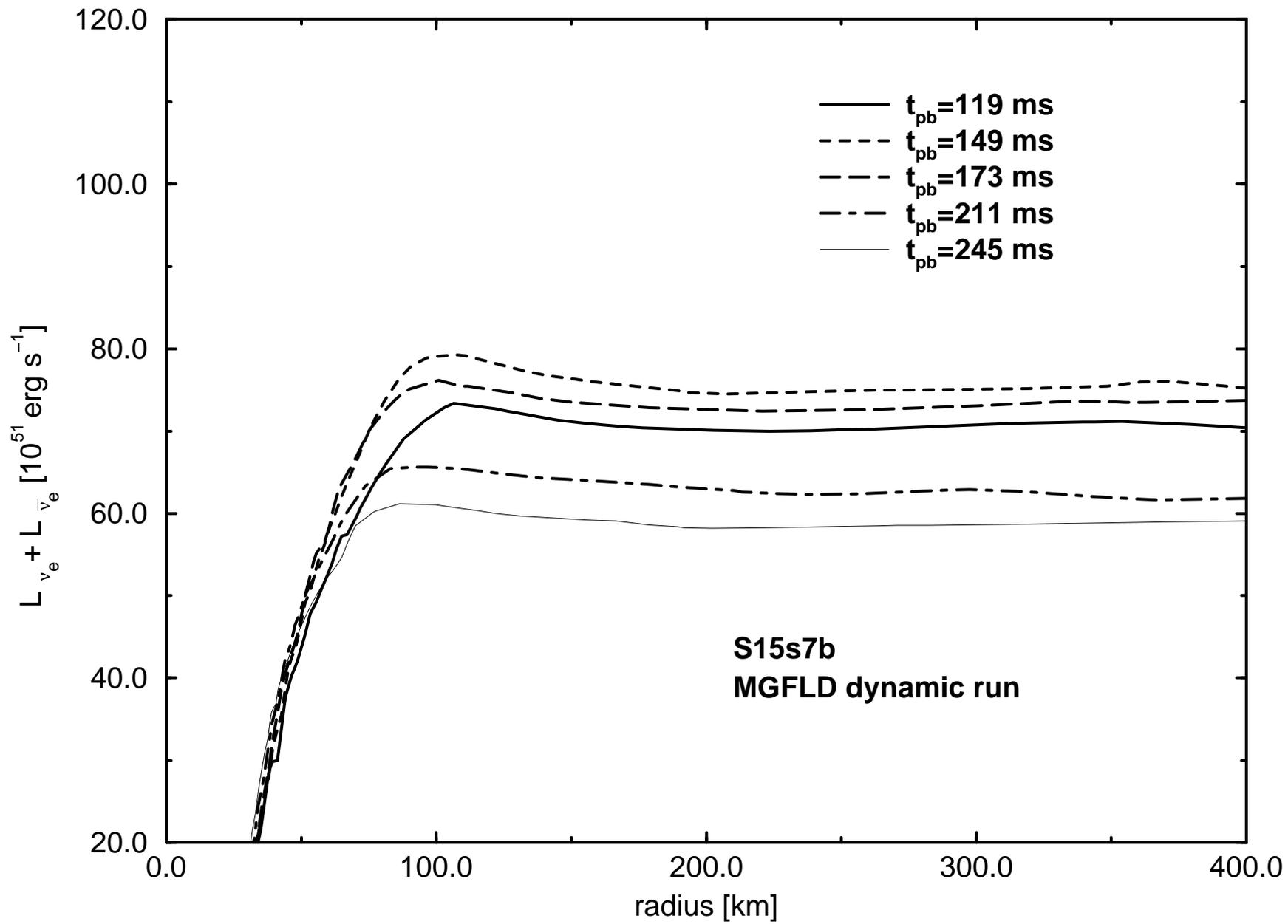
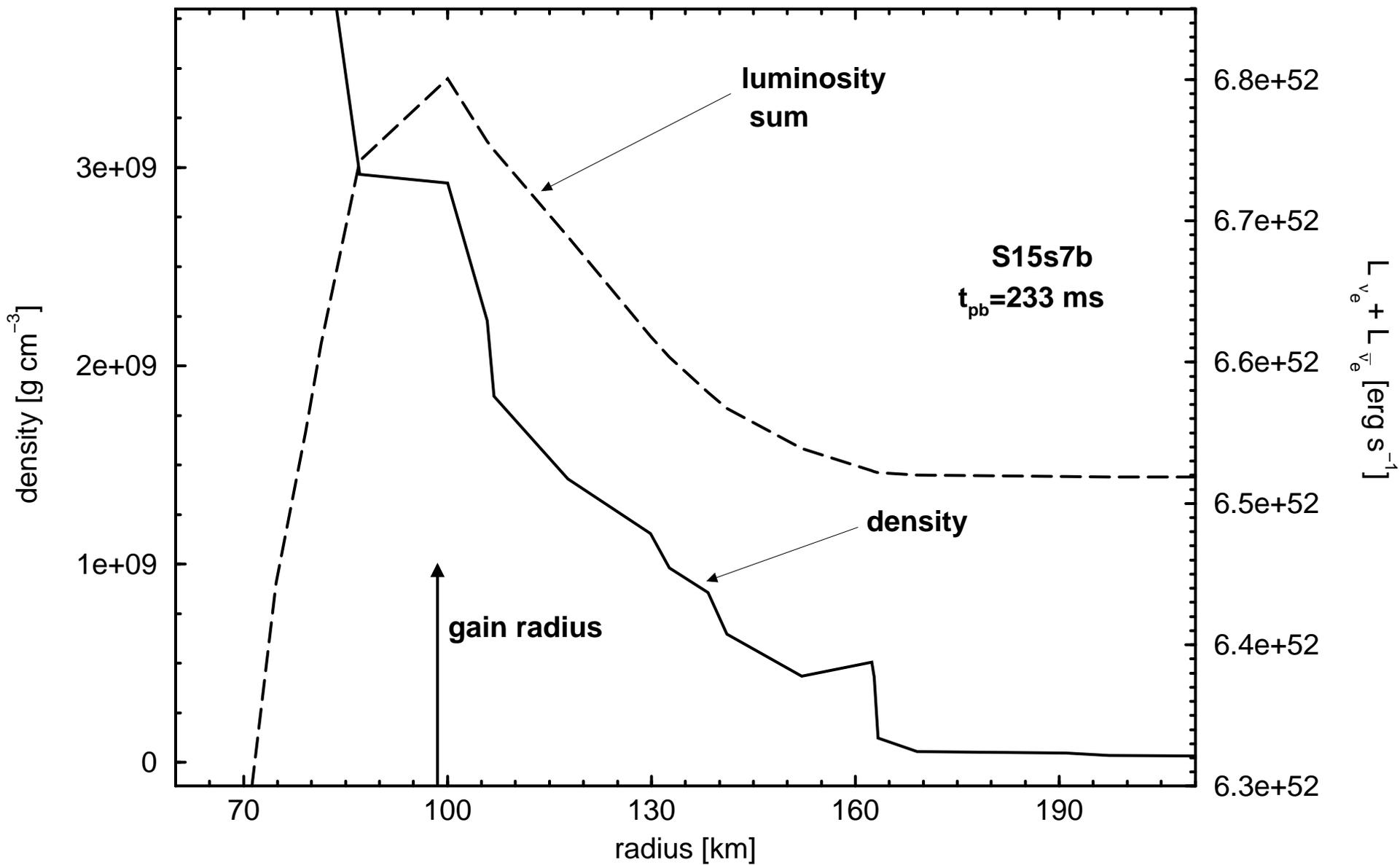
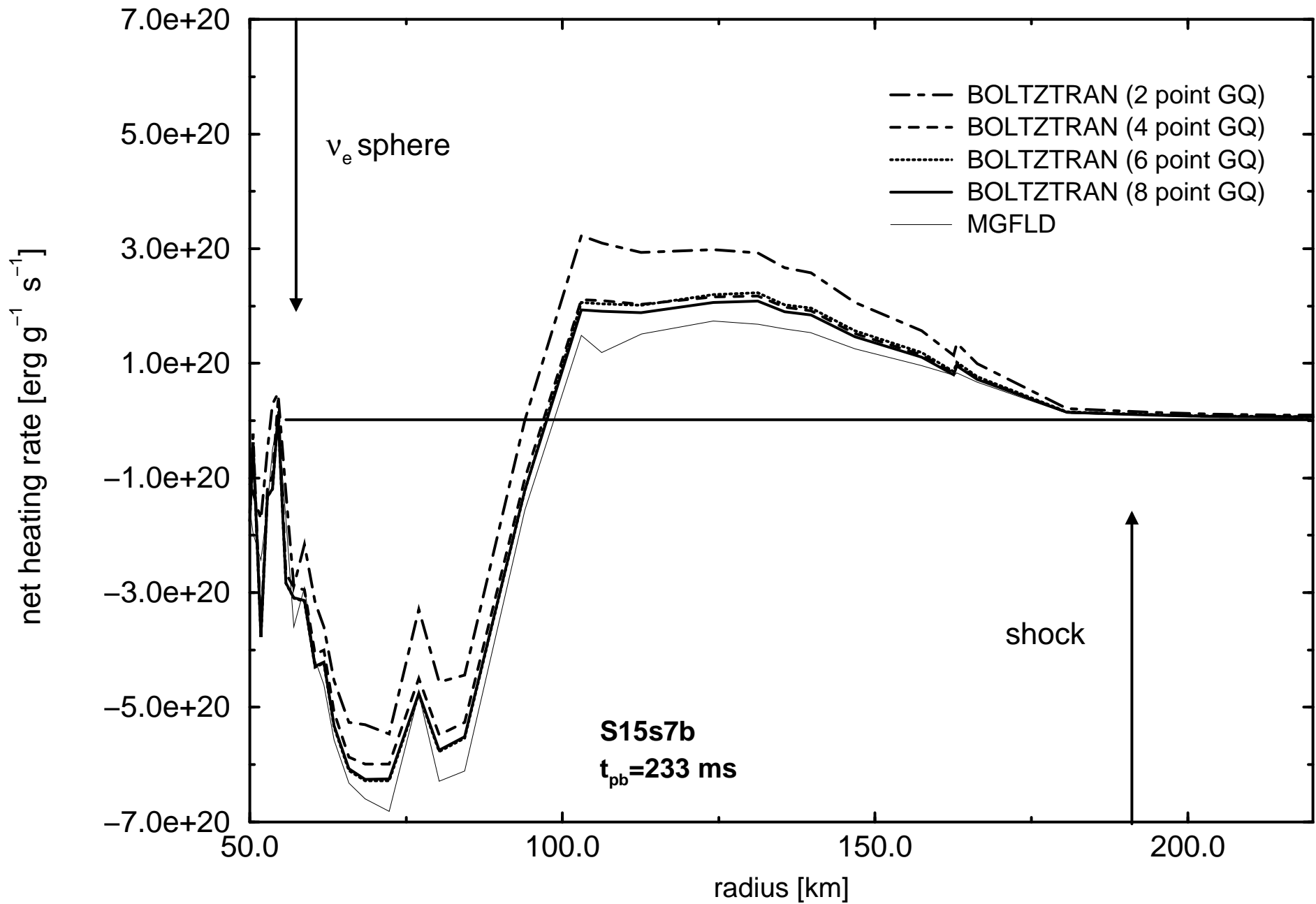
\end{document}